# BAYESIAN ESTIMATES OF THE LARGE-SCALE VELOCITY FIELD IN REAL SPACE AND REDSHIFT SPACE [1]

ALBERT STEBBINS
*NASA Fermilab Astrophysics Center, FNAL MS209, Box 500, Batavia Illinois 60510, USA.*

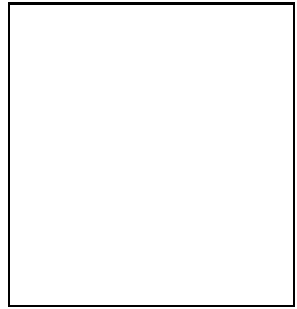

**Abstract**

Methods for inferring the velocity field from the peculiar velocity data are described and applied to old and newer data. Inhomogeneous Malmquist bias and ways to avoid it are discussed and utilized. We infer that these biases are probably important in interpreting the data.

## 1  Error and Bias

Measurements of large-scale peculiar velocities (LSPV) can provide very powerful tests of cosmological models, since they help us to determine the distribution of mass rather than the distribution of light. Bertschinger and Dekel [4] have shown that in spite of nonlinear clustering on small scales that one can use the velocity of matter on large scales to infer the large-scale mass distribution. However in using $D_n$-$\sigma$ and Tully-Fisher estimates of distances to infer the peculiar velocity field of galaxies one is beset with two major difficulties. The first difficulty is that the distance estimators are not very accurate, with 15%-20% fractional errors. Thus individual galaxies at moderate redshifts of $\sim 5000\,{\rm km/s}$ may have errors in estimates of their velocity which are much larger than the velocity one is trying to estimate. The second difficulty with inferring velocities from these inaccurate distance estimators is one of bias: that the errors may not only be large, but they may also lead to systematic over- or under-estimates of the velocity. Clearly both the error and the bias must be understood before one can proceed to apply this data to ruling in or out cosmological models on the basis of the LSPV data. In this contribution are presented methods for estimating velocities, errors, and biases which have been developed by Nick Kaiser and the author [12].

The canonical results on LSPV comes from applying the POTENT algorithm [7] to the distance estimator data to obtain an estimate of both the smoothed velocity field and the uncertainty in this reconstruction. The way that POTENT handles the large velocity errors is to average the estimated velocities of all the galaxies in a given region to estimate the average velocity in that region. In particular the galaxies' velocities are summed, weighted by a window function, to obtain an estimate of the smoothed velocity field. The Gaussian functional form of the window function is chosen for simplicity but the Gaussian width and galaxy weighting scheme is chosen using trial and error on Monte Carlo velocity fields. Estimates of the uncertainty and biases are also obtained from Monte Carlo simulations. The POTENT algorithm appeared to perform well in the Monte Carlo simulations of ref [7] however Monte Carlo simulations of POTENT by Newsam *et al.* [20] do indicate the existence of large biases depending on exactly which distance estimator is used. The question of statistical biases will be discussed more extensively below, but we will begin by suggesting an alternative method of assigning statistical weights.

---

[1] to appear in *Cosmic Velocity Fields* (Paris - July 1993) ed.s F. Bouchet and M. Lacheize-Rey

## 2 Regulated Inversion and Bayesian Estimators

There is nothing wrong with the trial and error method of assigning statistical weights (i.e. the window function) used by POTENT. However there are other methods of assigning statistical weights which are less *ad hoc* and which under certain circumstances may be considered "optimal". The general problem with trying to fit a curve or a velocity field model to noisy data is the conflict between trying to get the model to match the data while trying to avoid putting features in the model which are not justified by the data points. We would like to use a "non-parametric" model with a very large number of degrees-of-freedom (dof) which thus imposes no restrictive geometry on the model. With this many dof one may choose a model which passes through all of the data points, however since the noisy data will exhibit wild fluctuations many of which are not real this would be a poor reconstruction procedure. To obtain sensible answers one must "regulate" the reconstruction procedure in some way.

Unfortunately there is no *a priori* method of determining precisely how believable a given fluctuation in the data is. In order to regulate the reconstruction one must impose some preconception about what the velocity field should look like. This may be done, as in the POTENT method [7], by tuning ones reconstruction procedure to reproduce the velocity fields when the data is drawn from a CDM model, or more explicitly in the Bayesian approach described below. In any case as the data gets better the imposed preconception will play less of a role and the data more. The "direction" from which the inferred velocity field approaches the true velocity field as the data improves will depend on the imposed preconception.

One can formalize ones preconceptions in terms of a *prior* probability distribution, $p_i$, for the velocity field. If the velocity field from which the data were truly drawn from $p_i$ then Bayes theorem tells us that the *posterior* distribution is given by

$$p_p(v_r(\mathbf{x})|\text{data})\, dv_r(\mathbf{x}) = \frac{p_e(\text{data}|v_r(\mathbf{x}))}{p(\text{data})} p_i(v_r(\mathbf{x}))\, dv_r(\mathbf{x}) \;. \qquad (1)$$

where the $p$'s are functionals of the radial velocity field, $v_r(\mathbf{x})$; $p_e$ gives the assumed known distribution of the velocity estimator errors, and $p(\text{data})$ only plays the role of a normalizing constant. The 1st term of the rhs is the likelihood function for $v_r(\mathbf{x})$ which is larger for $v_r(\mathbf{x})$ which are better fits to the data and smaller for less good fits. The goodness-of-fit refers only to the smallness of the difference between the measured estimates of the peculiar velocities and the model velocities. Multiplying this is our prior distribution, $p_i$, which is large for what we might consider reasonable $v_r(\mathbf{x})$ and smaller for fields which are unreasonable. Thus $p_i$ plays the role of the regulating function, which is used to exclude wildly oscillating $v_r(\mathbf{x})$ which are nevertheless good fits to the data.

If one wanted to choose a particular realization of $v_r(\mathbf{x})$ as a guess for the true $v_r(\mathbf{x})$, a natural choice given the posterior distribution of eq 1, would be the expectation value for $v_r(\mathbf{x})$ under this distribution, i.e.

$$\bar{v}_r(\mathbf{x}) = \int v_r(\mathbf{x})\, p_p(v_r(\mathbf{x})|\text{data})\, dv_r(\mathbf{x}) \;. \qquad (2)$$

If the $p_i$ were truly the distribution from which $v_r(\mathbf{x})$ then clearly $\bar{v}_r(\mathbf{x})$ would be the unbiased estimator of $v_r(\mathbf{x})$. While $\bar{v}_r(\mathbf{x})$ may not oscillate wildly it will not be a realistic velocity field. One will find that at distances far from any velocity data that $\bar{v}_\mathrm{r} \to 0$. For most reasonable priors the sign of $v_\mathrm{r}$ at positions far from any data is just as likely to be positive or negative. Therefore $\bar{v}_\mathrm{r} = 0$ is a good estimator since it is, on average, closer to the truth than other values. We may also want to estimate the density or gravitational potential which generates these velocities, or their spatial averages. These quantities are, for small perturbations and irrotational flow, linearly related to $v_\mathrm{r}$, i.e $\Delta = \mathcal{L}(v_\mathrm{r}(\mathbf{x}))$. A good estimator of these quantities are their expectation values under the posterior distribution which is given by $\bar{\Delta} = \mathcal{L}(\bar{v}_\mathrm{r}(\mathbf{x}))$. This is what we will use below. From $p_\mathrm{p}$ one may also estimate how accurate these estimates are. The mean square deviation of the true value of $\Delta$ from its mean value estimate is given by

$$\overline{(\Delta - \bar{\Delta})^2} = \int \Big(\mathcal{L}(v_\mathrm{r}(\mathbf{x})) - \mathcal{L}(\bar{v}_\mathrm{r}(\mathbf{x}))\Big)^2 p_\mathrm{p}(v_\mathrm{r}(\mathbf{x})|\mathrm{data})\, \mathrm{d}v_\mathrm{r}(\mathbf{x}) \ . \tag{3}$$

In summary, if one is able to chose a prior distribution, $p_\mathrm{i}$ and able to perform the integrals of eq 2 & 3, then one has an estimator of most quantities of interest and an estimate of how accurate these estimators are. This method of assigning statistical weights may be used in place of that used in the POTENT [7] or Max-Flow [21] algorithms. All methods of assigning statistical weights have certain inherent biases in them. With the Bayesian method one chooses the biases explicitly.

While mentioning POTENT and Max-Flow we should include a word about sampling gradient bias and non-radial paths. The statistical weights that a Bayesian posterior distribution assigns to the galaxies are the correct ones for the sample of galaxies one uses and for the prior distribution one is assuming. Since the sampling is properly accounted for *there is no sampling gradient bias* as defined in ref [7]. Furthermore since one is not explicitly integrating along paths to find the velocity potential the question of integration along radial or non-radial paths never arises. The correct statistical weights are assigned by construction.

Of course the problem is in choosing $p_\mathrm{i}$. Almost any reasonable choice will serve the function of regulating the fitting procedure and thus preventing wild oscillations in $\bar{v}_\mathrm{r}$. If one chooses a $p_\mathrm{i}$ in which the data is very improbable then the $\bar{v}_\mathrm{r}(\mathbf{x})$ one will obtain will be an unhappy compromise in which neither goodness-of-fit to the data nor goodness-of-fit of $\bar{v}_\mathrm{r}(\mathbf{x})$ to the prior distribution are obtained. Unless ones prejudices are very strong one would probably like to only consider $p_\mathrm{i}$'s which yield $\bar{v}_\mathrm{r}(\mathbf{x})$'s which are a good fit to the data. One method of finding such $p_\mathrm{i}$'s in advance is to consider a class of models with a small number of parameters, say Gaussian random noise with variable amplitude and spectral index, and then use the likelihood function for these parameters to determine acceptable values. One would probably want to do this anyway since constraining model parameters is the main reason for looking at the peculiar velocities.

If one assumes Gaussian random noise for the prior distribution and assumes that the errors in the peculiar velocity estimators are Gaussian then the posterior distribution is also Gaussian. The distance estimator errors are usually assumed log-normal but since the fractional errors are small a Gaussian is not a bad approximation. If one has a set of estimators, $\{V_\mathrm{r}^a\}$, of the radial peculiar velocity at the set of positions, $\{\mathbf{x}_a\}$, then the expectation value of some field $\Delta(\mathbf{x})$ under the posterior distribution is given by [23, 14]

$$\overline{\Delta(\mathbf{x})} = \sum_a \sum_b \langle \Delta(\mathbf{x}) v_\mathrm{r}(\mathbf{x}_a) \rangle A_{ab}^{-1} V_\mathrm{r}^b \qquad \text{where} \qquad A_{ab} = \langle V_\mathrm{r}^a V_\mathrm{r}^b \rangle \tag{4}$$

where $\overline{\cdots}$ indicates an average under the posterior distribution and $\langle \cdots \rangle$ indicates an average under the prior distribution. Note that the difference between $\mathbf{v}_\mathrm{r}(\mathbf{x}_a)$ and $V_\mathrm{r}^a$ is that the latter includes the velocity estimator errors and the former does not. The $\langle \cdots \rangle$ are easy to compute in terms of the assumed power spectrum no matter whether $\Delta$ is density, velocity, or potential. The error is in the estimators, $\overline{\Delta(\mathbf{x})}$, are given by the correlation function

$$\overline{\left(\Delta(\mathbf{x}) - \overline{\Delta(\mathbf{x})}\right)\left(\Delta(\mathbf{y}) - \overline{\Delta(\mathbf{y})}\right)} = \langle \Delta(\mathbf{x})\Delta(\mathbf{y}) \rangle - \sum_a \sum_b \langle \Delta(\mathbf{x}) v_\mathrm{r}(\mathbf{x}^a) \rangle A_{ab}^{-1} \langle \Delta(\mathbf{y}) v_\mathrm{r}(\mathbf{x}^b) \rangle \ . \tag{5}$$

Since the distribution is Gaussian the mean value of eq 4 and the correlation function of eq 5 fully determine the entire distribution. Even if the distribution is not expected to be Gaussian the mathematical convenience of being able to sum over all realizations analytically argues for using the Gaussian assumption, at least as a first approximation. The use of a Gaussian prior does not exclude $\overline{\Delta(\mathbf{x})}$ from having "non-Gaussian" behavior. The random-phase property of Gaussian distributions means that the Gaussian prior has no preference for any particular phase-correlations but the posterior distribution may have phase-correlations if the data prefer it.

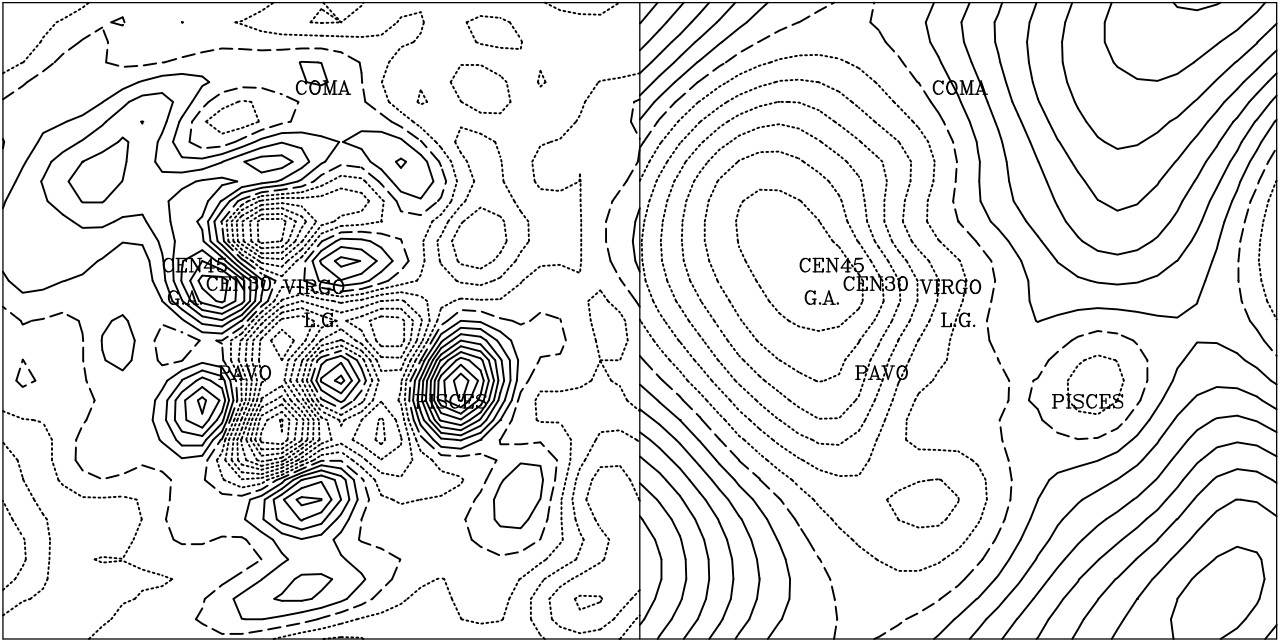

**FIGURE 1** Shown are the Bayesian reconstructions in the Supergalactic Plane of the overdensity, $\delta\rho/\rho$, (left) and the gravitational potential, $\Phi$ (right). The solid lines are positive contours, the dotted lines are negative contours, and the dashed line the zero contour. The contours are in units of 0.2 for $\delta\rho/\rho$ and $10^{-5}/c^2$ for $\Phi$. The locations of several galaxy clusters, the Local Group, and the Great Attractor [17] are indicated. The plotted region is $200 h^{-1}$Mpc on a side. The prior distribution is Gaussian with an $n = -1$ power-law power spectrum an amplitude corresponding to a 1000 km/s 3-d velocity dispersion. The power is cutoff with an $18\, h^{-1}$Mpc resolution on small scales and a $200\, h^{-1}$Mpc resolution on large scales (hence the periodicity). The galaxies are positioned according to their estimated distances.

## 3 Bayesian Reconstruction in Real Space

In figs 1-3 are reconstructions of the overdensity field and the gravitational potential field obtained by the Bayesian technique using 2425 estimates of peculiar velocities provided by D. Burstein (Mark II release) [1, 2, 3, 5, 8, 9, 16, 17], J. Willick [24], and D. Mathewson [18, 19]. For the Mark II release we use Burstein's estimate of the velocity-estimator errors. We have used the homogeneous Malmquist corrected velocities and where group assignments are provided we have combined all the galaxies in the group into one data point. This should reduce the inhomogeneous Malmquist bias in both real-space and redshift-space. The reconstruction was done in a cubical box, $L = 200\, h^{-1}$Mpc on a side, aligned with the Supergalactic coordinate system, and centered on the Local Group. Only measured galaxies within this box are included, the galaxies are placed at their homogeneous Malmquist corrected estimated distances, and the velocities used are in the Cosmic Microwave Background (CMB) frame. The prior distribution is Gaussian with a power-law power spectrum and amplitude corresponding to an expected 1000 km/s 3-d rms velocity for each galaxy. However the spectrum only applies to modes inside the box and only modes with wavenumber $k \leq 14\pi/L$ are included, corresponding to a $18\, h^{-1}$Mpc resolution limit. To account for velocities generated on scales below our resolution we add a $(250\,\text{km/s})^2$ 1-d "field dispersion" to each galaxy and group. This field dispersion of each group is assumed statistically independent for each galaxy and is included in quadrature in the 1000 km/s.

The spectral indices chosen are $n = -1$, $n = +1$, and $n = -3$. An $n = +1$ spectrum expects most

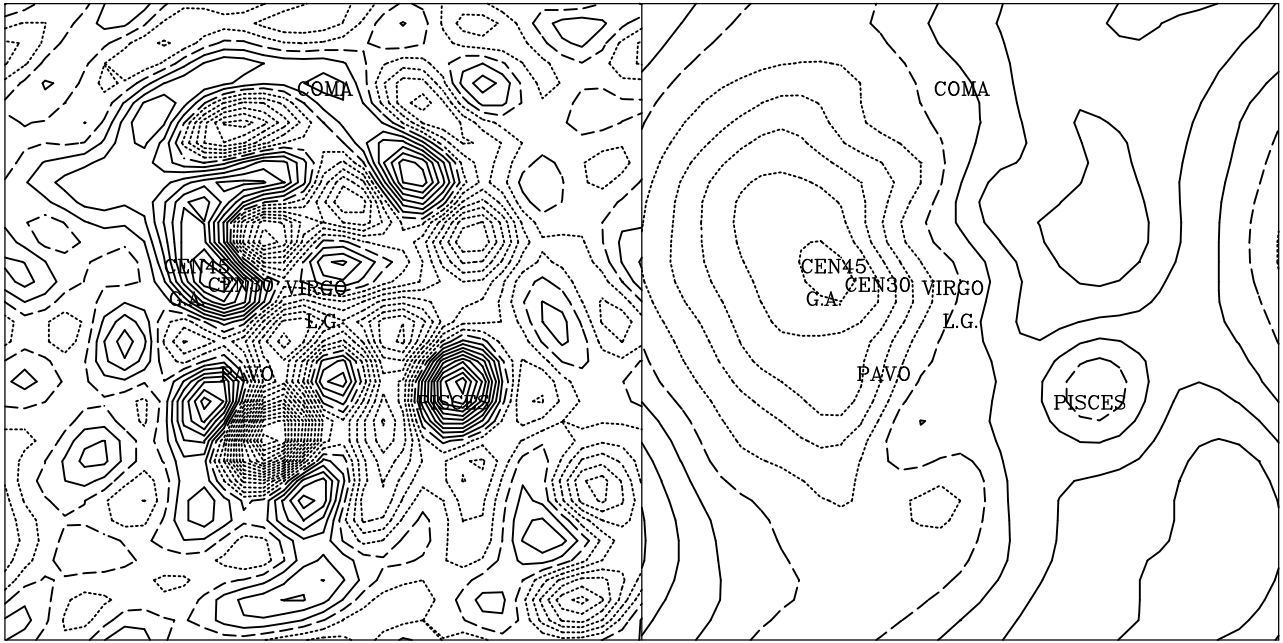
**FIGURE 2** Same as for figure 1 except $n = +1$.

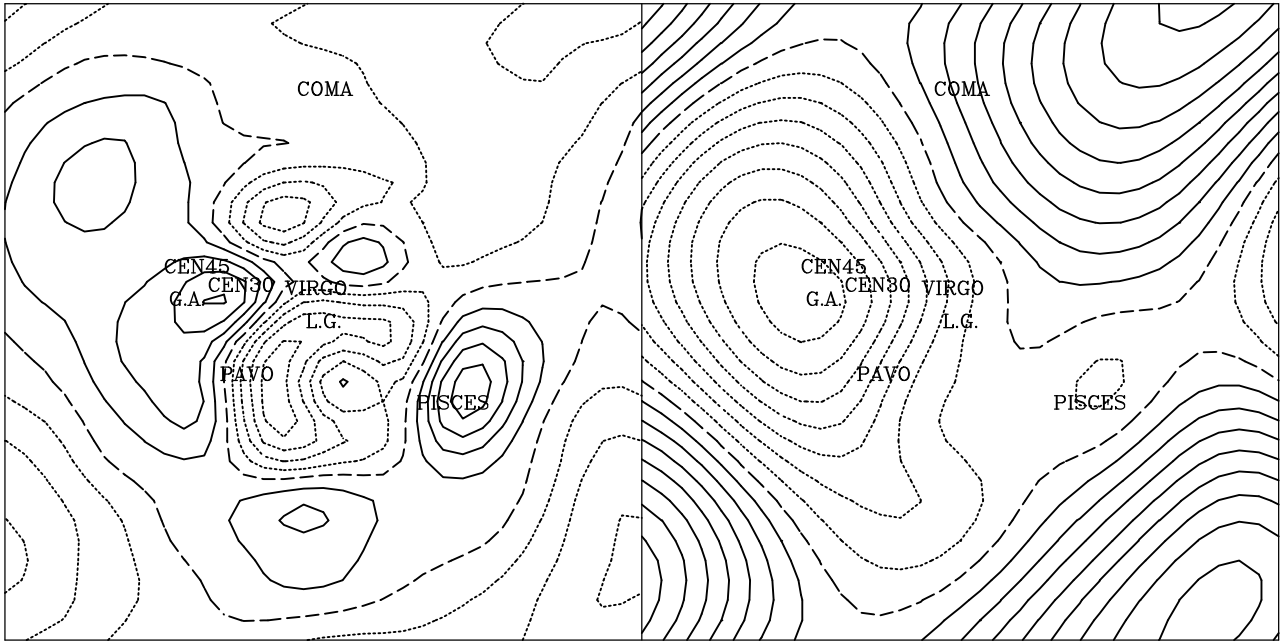
**FIGURE 3** Same as for figure 1 except $n = -3$.

of the velocity to be generated on small scales near the resolution scale, an $n = -3$ spectrum expects most velocities generated on large scales near the box size and an $n = -1$ spectrum expects velocities generated equally on all scales. On these scales a spectral index of $\approx -1$ is what one might expect from the galaxy distribution [11, 10]. The goodness-of-fit of all three of these model velocity fields is about the same: a reduced $\chi^2$ of $\approx 1.7$ for 2425 dof. This is, of course, not a very good fit. A more generous field dispersion would have made this smaller. Since these models have roughly the same goodness-of-fit there is no strong reason to choose one over the other.

We see that by comparison the density fields under these three priors are significantly different. As one would expect, for the $n = +1$ model there is much more substructure, somewhat less in the $n = -1$ model, and much less in the $n = -3$ model. In contrast the general structure of the potential field is rather similar between the different models. There is a deep potential well with bottom near the Great Attractor [17] and a much weaker well near Pisces. The Great Attractor shows no evidence of being a monolithic mass concentration, with nearby voids in all models. However the overall picture for the flow, including mass concentrations both above and below the plane, seems to be infall into the Great Attractor. Unfortunately we will argue that many of the structures seen here are artifacts.

## 4 Inhomogeneous Malmquist Bias

In deriving the Bayesian method described above it has been assumed that the velocity estimators used are drawn from a probability distribution with the properties:
**I)** the mean value is equal to the true velocity at the position assigned to the galaxy,
**II)** the variance is known,
**III)** the error distribution is statistically independent for each galaxy, and
**IV)** the distribution is Gaussian.
A first step toward satisfying **I)** is to obtain an unbiased "raw" distance estimator: $\hat{d}$, one whose expected value would give the correct distance (and hence peculiar velocity) for a given galaxy. It is argued [22, 13] that the raw distance estimators used so far may be biased due to the way the galaxies are selected and other distance estimators have been proposed. Of course, what one really needs is an unbiased estimator of the velocity at a given position which is not usually given by an unbiased raw distance estimator. The main difficulty with achieving this goal is that we have no way of knowing the true distance to a given galaxy.

What is usually settled for (c.f. in the above analysis) is the peculiar velocity as a function of estimated position. While an unbiased raw distance estimator guarantees that the amount of distance overestimation is balanced by the amount of distance underestimation when averaging over galaxies at a fixed distance, it does not guarantee that this is true for all galaxies at a fixed estimated distance. This asymmetry can occur when
**1)** the distribution of distance estimators is not symmetric about the true distance, and/or
**2)** the distribution of galaxies at larger and smaller distances is not the same because
 **a)** there is more volume and hence more galaxies at larger distances,
 **b)** the galaxies are not uniformly distributed in space,
 **c)** the galaxy samples used are not uniformly selected in space.
In order to avoid spurious peculiar velocities caused by these effects one should apply a "Malmquist" correction to the distance estimator: $\hat{d} \rightarrow \hat{d}_M$. The Malmquist correction is the offset required to assure that the average estimated distance equals the average true distance of all the galaxies at a given estimated distance and similarly for the average peculiar velocities. One then assigns the galaxies positions at a distance, $\hat{d}_M$, and estimates the peculiar velocity by $\hat{v}_M = cz - H_0 \hat{d}_M$. Since the size of the biases from **2b)** and **2c)** are often difficult to estimate usually only a "homogeneous Malmquist correction" (see [17]), which accounts for only **1)** and **2a)**, is used. Leaving aside selection effects, one could rightly argue that the homogeneous correction is the proper correction, when averaging over different realizations of the galaxy distribution. One could properly use such a homogeneous correction if one were prepared to take into account the strong correlations in the distance errors between neighboring galaxies the inhomogeneous galaxy distribution introduces. These correlations would strongly violate assumption **III)** and would also require modeling of the statistics of the galaxy distribution. Landy and Szalay [15] have proposed a technique for estimating the "inhomogeneous" bias due both to **2b)** and **2c)**, which uses only the galaxies whose distances are estimated. While these inhomogeneous corrections have not been much used so far, it is clear from the results of refs [15, 20] that they make significant corrections to the inferred flows, especially at the edge of ones sample.

**FIGURE 4** Illustration of inhomogeneous Malmquist bias: **a)** $\delta(d)$ (*solid*), $\delta_g(d)$ (*dotted*), and $\hat{\delta}(\hat{d})$ (*dashed*) derived from $\bar{z}(\hat{d})$, **b)** $z(d)$ (*solid*) and $\bar{z}(\hat{d})$ (*dashed*) derived from the $\hat{d}$-$z$ galaxy distribution, **c)** the $\hat{d}$-$z$ distribution of galaxies assuming 15% distance errors. N.B. The bias is the exaggeration of the features in the solid curves by the dashed curves.

To illustrate how this inhomogeneous bias works we present the results from a 1-d Zel'dovich simulation in fig 4. The solid line in the 1st panel of shows the density vs. distance from the observer of a mildly nonlinear density perturbation. The galaxy number density illustrated by the dotted line follows the mass density, but with a bias factor of 2. The mass inhomogeneities produce peculiar velocities which give the redshift-distance relationship shown by the solid line in the 2nd panel. This exhibits an s-wave pattern of infall onto the positive mass concentration and an anti-s-wave outflow around the negative mass concentration. In 1-d one can infer the mass overdensity from this curve via

$$\delta = \frac{1}{f(\Omega_0)}\left(1 - \frac{dz}{dd}\right) \qquad f(\Omega_0) \approx \Omega_0^{0.6} \qquad (6)$$

In the 3rd panel we have drawn 2000 galaxies from our distribution and scattered them in $\hat{d}$-$z$ space using log-normal distance errors, 15% scatter, and the homogeneous Malmquist corrections [17]. The usual procedure is to take a distribution, average at a fixed $\hat{d}$, to obtain $\bar{z}(\hat{d})$. This function is then used to infer velocities and densities. The dashed line in the 2nd panel gives the $\bar{z}(\hat{d})$ one would obtain in the limit of infinite sampling. The central horizontal bar comes from the concentration of galaxies near the center. We see that these galaxies tends to draw up $\bar{z}(\hat{d})$ in front of the concentration and to draw down $\bar{z}(\hat{d})$ in the back. The opposite effect occurs for the more distant void. The net effect is to over-accentuate the the s-wave and anti-s-wave. The overdensity one would infer by using $\bar{z}(\hat{d})$ in eq 6 is given by the dashed line in the 1st panel. We see that the masses of features are overestimated.

To get a feel for the size of this effect consider a small amplitude 1-d sinusoidal perturbation along the line-of-sight with a constant galaxy bias, $b$, and a normally distributed distance error with *constant* rms, $\sigma$. In this case the homogeneous Malmquist correction is zero. In the limit of infinite sampling and small amplitude the inferred mass density and velocity as a function of $\hat{d}$ is related to the true density and velocity in real space by

$$\hat{\delta}_g(\hat{d}) = e^{-\frac{1}{2}(k\sigma)^2}\delta_g(d) \qquad \hat{v}(\hat{d}) = \left(1 + \frac{b}{f(\Omega_0)}(k\sigma)^2\right)e^{-\frac{1}{2}(k\sigma)^2}v(d) \qquad (7)$$

where $k$ is the wavenumber. If one were to infer $f/b$ in the usual way one would obtain

$$\frac{\widehat{f(\Omega_0)}}{b} \equiv \frac{-\nabla_{\hat{d}}\hat{v}}{\hat{\delta}_g} = \left(\frac{f(\Omega_0)}{b} + \pi^2\left(\frac{\sigma}{\lambda/2}\right)^2\right) \qquad \lambda = \frac{2\pi}{k} \ . \qquad (8)$$

The 2nd term in this equation is present whether or not there are any true velocities. Let us apply this simplistic model to a simplistic model of our own neighborhood: we live in a trough of galaxies between the large concentrations of galaxies in Hydra-Centaurus and Perseus-Pisces. The peak-to-peak wavelength is $\lambda \sim 100\,h^{-1}$Mpc and taking 15% of half this distance for $\sigma$ one finds a spurious addition to $f/b$ of 0.22 which could add significantly to $\Omega_0$ if the true value is near 1. Better estimates of these biases can and should be gotten by applying the inhomogeneous Malmquist corrections to the data.

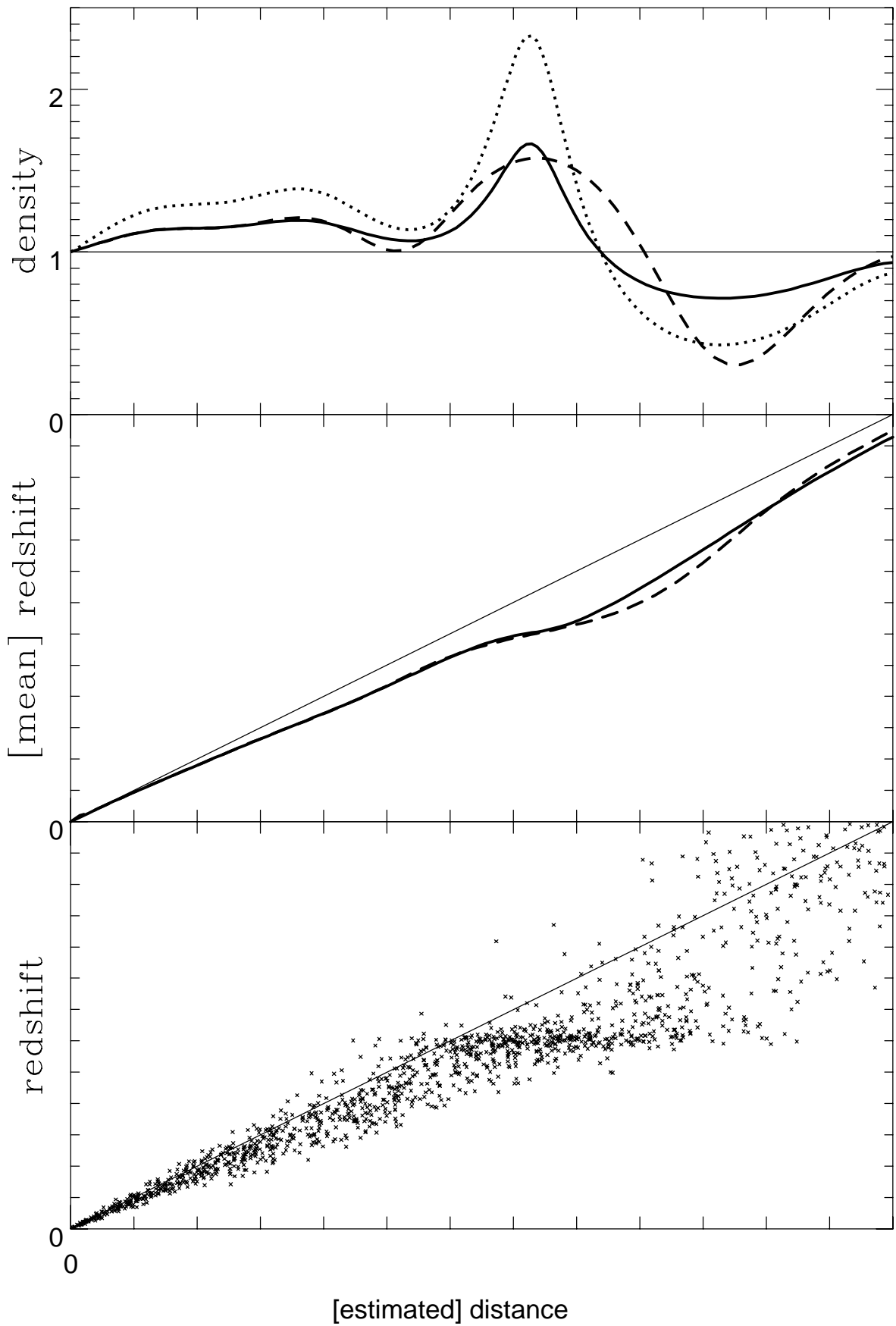

[estimated] distance

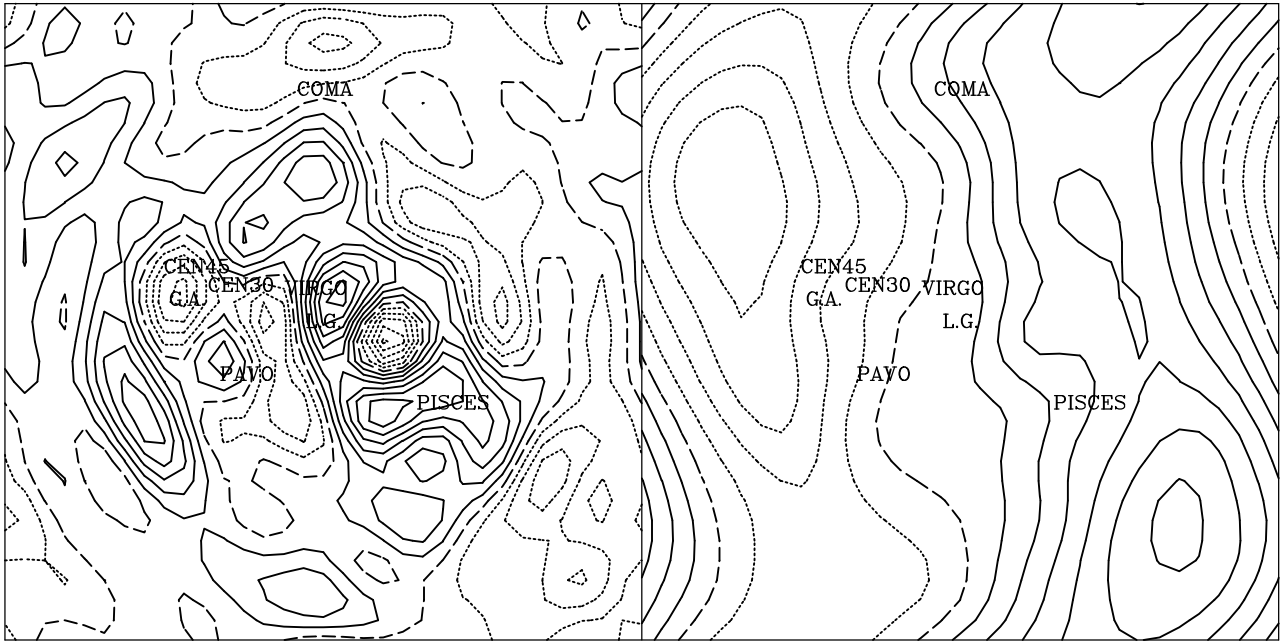

**FIGURE 5** Same as for figure 1 except the galaxies are positioned at their redshift distances and no homogeneous Malmquist correction is applied.

## 5 Bayesian Reconstruction in Redshift Space

Instead of the traditional approach outlined above one can instead try to estimate the velocity field in redshift space which avoids much of the problems with Malmquist corrections. If one imagines that the flow is cold, i.e. there is a well defined velocity at each position with negligible dispersion, then all of the galaxies at a given redshift and position on the sky are at the same position in space. Thus if one has an unbiased raw distance estimator it will yield an unbiased estimator of distance and peculiar velocity when averaging over galaxies at a given position in redshift space. The velocity field thus obtained is not biased by galaxy clustering or selection effects the way it is in estimated distance space. Furthermore since redshifts can be measured accurately there is not the same ambiguity of where to place the galaxy. If one is able to obtain an accurate redshift space velocity field then one can easily transform this into the real space velocity field. Of course the cold-flow approximation is not extremely accurate, pairwise velocity dispersions of $\sim 400$ km/s are found on fairly small scales [6] and much larger dispersions are found in clusters of galaxies. Clusters are fairly easy to identify on the sky and if one combines the galaxies in a cluster into a single object one can decrease the effective dispersion considerably. The effect of the remaining dispersion is similar to the usual inhomogeneous Malmquist effect however it has just the opposite sign: it creates false infall into voids and false outflow out of galaxy concentrations thus underestimating the density contrasts. Redshift is a more accurate distance indicator than the traditional $D_n$-$\sigma$ or Tully-Fisher at the distances of the Great Attractor or Perseus-Pisces. The redshift space Malmquist effects are thus smaller than those in estimated distance space.

Figure 5 presents the results of an $n = -1$ velocity reconstruction, performed as before except using redshift distances and using velocity estimators which are not Malmquist corrected. The small redshift to real space conversion has not been performed which may leave noticeable artifacts in the density field. The most striking difference between this reconstruction and the previous ones, is that the main potential well has moved beyond the Great Attractor, indicating flow past it. The fact that the well is in the box at all is partly due to periodicity in a finite box. It would be difficult to determine where the flow converges since there are very few galaxies at these distances. The reason for the difference between this reconstruction and the previous one is that the average raw velocity estimators at a given redshift do not show the convergence of flow that is exhibited by the homogeneous Malmquist corrected velocity estimators averaged at a given estimated distance. The two flows should be similar if slightly distorted versions of each other. Since this is not the case it seems likely that the traditional method is significantly contaminated by inhomogeneous Malmquist bias. Fig 5 is probably closer to the truth.

**Acknowledgements.** Thanks to Dave Burstein, Jeff Willick, and Don Mathewson for providing their data. Thanks also to Nick Kaiser with whom much of the work reviewed here has been done. AS was supported in part by the DOE and the NASA through grant number NAGW-2381.